\documentclass{elsarticle} 

\usepackage{framed}
\usepackage{enumitem}

\usepackage{times}

\usepackage[T1]{fontenc}
\usepackage[scaled=0.81]{beramono}
\usepackage{hyperref}

\newcommand{\e}[1]{\mbox{\lstinline[basicstyle=\normalsize]|#1|}}
\newcommand{\es}[1]{\mbox{\lstinline[basicstyle=\scriptsize]|#1|}}

\usepackage{amsmath,amssymb}
\usepackage{xcolor}
\usepackage{booktabs}
\usepackage{colortbl}
\usepackage{xspace}

\usepackage{listings}
\usepackage{eiffel}
\usepackage{boogie}

\usepackage{zed-csp}

\begin{document}

\title{Seamless Requirements}

\author[an]{A.~Naumchev}
\ead{a.naumchev@innopolis.ru}

\author[an,bm]{B.~Meyer}
\ead{Bertrand.Meyer@inf.ethz.ch}

\address[an]{Innopolis University, Innopolis, Russia}
\address[bm]{Politecnico di Milano, Milan, Italy}

\begin{abstract}
Popular notations for functional requirements specifications frequently ignore developers' needs, target specific development models, or require translation of requirements into tests for verification; the results can give out-of-sync or downright incompatible artifacts. Seamless Requirements, a new approach to specifying functional requirements, contributes to developers' understanding of requirements and to software quality regardless of the process, while the process itself becomes lighter due to the absence of tests in the presence of formal verification. A development case illustrates these benefits, and a discussion compares seamless requirements to other approaches.
\end{abstract}


\maketitle

\section{Introduction} \label{sec:intro}

Seamless Requirements is a technique to close the various gaps that have long plagued the practice of software requirements:

\begin{itemize}
  \item The gap between customers and developers (\autoref{sec:sec:intro_reuse_by_stakeholders}).
  \item The gap between agile and formal development (\autoref{sec:sec:intro_reuse_by_models}).
  \item The gap between construction and verification (\autoref{sec:sec:intro_reuse_by_activities}).
\end{itemize}

To reach this goal, seamless requirements build on ideas coming from diverse sources, including literate programming \cite{knuth1984literate}, multirequirements \cite{Meyer13Multi}, and formal verification \cite{tschannen2015autoproof}. A seamless requirement combines two elements: a contracted self-contained routine, which doubles as a proof obligation, and an associated natural language comment.

The approach assumes object-oriented non-concurrent setting and does not handle non-functional requirements.

\subsection{Customers vs. Developers} \label{sec:sec:intro_reuse_by_stakeholders}

By adding programming languages with contracts to the family of requirements specification notations, seamless requirements improve developers' understanding of requirements that typically exist in some declarative form that has nothing to do with programming.

The modern taxonomy of requirements specification languages (\cite[Chapter~4 ``Requirements Specification and Documentation'']{van2009requirements}) provides a number of formal and semi-formal notations, and programming languages are not a part of this taxonomy. This implicitly isolates people (customers) who state requirements from people (developers) who implement them. As soon as the customers elicit and document requirements, demonstrate some ``good'' properties of the requirements within the chosen notation, the developers will have to map the notation into the semantics of the target programming language. Is there any way to check the translation at the same level of rigor used to derive those ``good'' properties of the requirements? Some approaches advocate modeling software at different angles using different notations to ensure its proper understanding by developers, but such an approach raises the problem of potential inconsistency between the views.

Seamless requirements express software functionality using the language best understood by developers: the programming language. The idea is not new \cite{Meyer13Multi}, but its implementation is (\autoref{sec:sec:multirequirements} gives more details). A seamless requirement is a compilable contracted self-contained routine - specification driver \cite{anaumchev2016drivers} - equipped with a structured natural-language comment. The comment delivers the meaning of the requirement to the customers, and the program construct - to the developers. Specification drivers are expressive enough to fully capture algebraic specifications \cite{anaumchev2016drivers}, and exercising their expressiveness is a driving force of the present research.

The idea of combining formal and natural language descriptions is present in goal-oriented requirements engineering \cite{van2001goal}, but the approach does not consider a programming language as a formal notation.

\subsection{Agile vs. Formal Development} \label{sec:sec:intro_reuse_by_models}

By nature both self-contained and formal, seamless requirements boost reliability of software produced using agile processes.

Compatibility of agile development and formal methods has long been a concern for software engineers (\cite{turk2014limitations}, \cite{black2009formal}), particularly in the domain of mission and life-critical systems (\cite{drobka2004piloting}, \cite{sidky2007determining}). The studies have something in common: their main concern is integration of agile practices into development of software that has to be reliable and is currently developed using some conservative process. In the same time there is a lack of research that studies applicability of formal methods to agile development of not so critical mass-market software for increasing its reliability. This problem is among the concerns of the present article.

In agile development a functional requirement typically takes a form of a scenario describing user interaction with the to-be software. The scenario is then translated into a set of unit tests for ensuring functional correctness of the software with respect to the scenario. Scenarios and unit tests naturally fit the agile philosophy of frequently delivering software in small increments: they both are self-contained information units suitable for grouping into arbitrary sized sets. It is the very nature of tests that limits the level of formality in agile development: they exercise only a subset of the possible execution paths. Although there are scientific approaches for making a test suit cover the software well enough, agile methods do not consider tests as a very important artifact and do not advocate improving tests coverage too much.

Seamless requirements replace unit tests with specification drivers, testing with formal verification, and move structured natural language scenarios to comments on specification drivers. Specification drivers can capture scenarios in their abstract form (as opposed to unit tests), which is why it makes sense to conjoin them. The resulting requirement form keeps the fine granularity of tests and scenarios, while being mathematically formal.

\subsection{Construction vs. Verification} \label{sec:sec:intro_reuse_by_activities}

Seamless requirements enable straightforward verification of existing software with respect to requirements without introduction of intermediate artifacts such as tests.

The modern software mass market rests on testing as the primary mechanism for checking functional correctness. Although tests are fundamentally imprecise (\autoref{sec:sec:intro_reuse_by_models}), there are scientific approaches to testing that enable production of test suits having reasonable code coverage with respect to some predefined criteria \cite{jorgensen2016software}. Such an approach may be suitable for greenfield software construction, but not always for verification of existing software that already works somehow. The problem is real: software quality cannot be higher than that of its least quality component. This means that, in order to reuse a third-party component, the development team has to make sure that its quality conforms to the quality standards defined in the project; in particular, it is necessary to generate sufficient number of tests and run them on the component. It is not surprising that such an effort is often considered as waste: why test something that is already on the market and works instead of putting more effort into construction?

Seamless requirements fix the issue by replacing testing with formal verification of specification drivers, which are formal and abstract representations of software usage scenarios. The only assumption upon which the approach rests is existence of a contract in the component, which is dictated by modularity of the verification approach\cite{tschannen2015autoproof}.

\section{Motivating Example} \label{sec:problem}

A simple example illustrates the idea of seamless requirements. The task is to implement a clock class that features seconds, minutes, hours, and days of week. The clock state should be updated through a special command called "tick" that advances the seconds counter. There is also an existing class \e{CLOCK} that does not feature the current day of week. The class is implemented and specified in Eiffel \cite{meyer1988eiffel}. The implementation is closed: only a specification in the form of a contract is available. \autoref{fig:verifiable_clock_implementation} contains the visible part of the class. The \e{frozen} specifier prohibits inheriting from the \e{CLOCK} class and thus makes it usable only for instantiating and using its instances. It is also known that the hidden implementation of the \e{tick} command is provably correct with respect to its postcondition in \autoref{fig:verifiable_clock_implementation}, and the correctness was established with the AutoProof verifier \cite{tschannen2015autoproof} for Eiffel programs with contracts.

\begin{figure}
\begin{lstlisting}
frozen class CLOCK
feature
  second, minute, hour: INTEGER

  tick
    do
      -- Hidden implementation
    ensure
      old second <= 58 implies ((second - old second = 1) and minute = old minute)
      old second > 58 implies (second = 0 and (old minute <= 58 implies minute - 1 = old minute)
      	and (old minute > 58 implies (minute = 0 and (old hour <= 22 implies hour - old hour = 1)
      		and (old hour > 22 implies hour = 0))))
    end
end
\end{lstlisting}
\caption{The existing clock class} \label{fig:verifiable_clock_implementation}
\end{figure}

\subsection{Existing Code} \label{sec:sec:clock}

This section takes a closer look at the visible parts of the \e{CLOCK} class in \autoref{fig:verifiable_clock_implementation}. The space between the \e{do} and \e{ensure} keywords of the \e{tick} feature would typically contain executable instructions, which are hidden in this case. Logical assertions between the \e{ensure} and the closest \e{end} keyword constitute the postcondition of the feature. The postcondition logically connects the clock pre-state, which precedes any invocation of \e{tick}, with the post-state, which results from the invocation. The \e{old} keyword before some identifiers denotes values of the respective queries in pre-states. Accordingly, identifiers that go without the \e{old} keyword denote values of the respective queries in post-states.

Since it is not possible to modify the \e{CLOCK} class neither directly nor through inheritance, it seems reasonable in this context to implement the required extended class through the composition relation: in the new class declare a reference to an object of type \e{CLOCK} and reuse its functionality. In order
to do so it is necessary to make sure that objects of the existing class, indeed, behave like a real clock. The extended clock development plan thus consists of the following major steps:

\begin{enumerate}
  \item Identifying requirements to an extended clock.
  \item Identifying requirements that are applicable to a non-extended clock.
  \item Verifying the existing \e{CLOCK} class with respect to the requirements for a non-extended clock.
  \item Reusing the existing class in the event of its successful verification.
  \item Developing a completely new class otherwise.
\end{enumerate}

\subsection{Natural-language Requirements} \label{sec:sec:requirements}
Implementation of the first two steps of the plan starts with enumeration of the requirements in their natural-language form in \autoref{fig:natural_clock_requirements}. Requirements (REQ1)-(REQ8) do not talk about the current day of week and thus are applicable to the existing implementation. Requirements (REQ9)-(REQ11) talk about the days counter and thus are applicable only to the extended implementation. For simplicity, days are represented with numbers from 0 to 6, where 0 corresponds to Monday, and 6 - to Sunday.

In many cases natural-language requirements are less clear and precise than the ones in \autoref{fig:natural_clock_requirements}. This particular issue is irrelevant to the present work, which is why the example relies on the assumption that the natural-language requirements in the clock example are of high enough quality.

Step 3 of the plan from \autoref{sec:sec:clock} is to check whether the \e{CLOCK} class meets requirements (REQ1)-(REQ8). This step, along with steps 4 and 5, is far less trivial than steps 1 and 2 and raises a number of questions.

\begin{figure}
A clock tick:
\begin{description}
  \item [(REQ1)] Increments current second if it is smaller than 59.
  \item [(REQ2)] Resets current second to 0 if it equals 59.
  \item [(REQ3)] Increments current minute if the time is HH:MM:59 for MM smaller than 59.
  \item [(REQ4)] Resets current minute to 0 if it equals 59 and current second equals 59.
  \item [(REQ5)] Keeps current minute if current second is smaller than 59.
  \item [(REQ6)] Increments current hour if the time is HH:59:59 for HH smaller than 23.
  \item [(REQ7)] Resets current hour to 0 if the time is 23:59:59.
  \item [(REQ8)] Keeps current hour if current second is smaller than 59.
  \item [(REQ9)] Increments current day at 23:59:59 if it is not Sunday.
  \item [(REQ10)] Resets current day to Monday after a clock tick at 23:59:59 on Sunday.
  \item [(REQ11)] Keeps current day if current second is smaller than 59.
\end{description}
\caption{Natural-language requirements to clock} \label{fig:natural_clock_requirements}
\end{figure}

\subsection{Research Questions} \label{sec:sec:research_questions}

\subsubsection{\textbf{RQ1}} \label{sec:sec:sec:rq1}
\textit{How to express precise semantics of the natural-language scenarios (REQ1)-(REQ8) using programming language constructs?}

Natural-language statements in \autoref{fig:natural_clock_requirements} are comfortable for reading by human beings. This may be not enough, however, for those who will potentially implement the requirements. Natural language is a source of misinterpretations and ambiguities, which is why it is not enough to have
requirements in this form \cite{meyer1993formalism}. What do statements (REQ1)-(REQ8) mean exactly in terms of the programming language abstractions? It would benefit the software developers to be able to precisely express the requirements in the programming language that will later be used for their implementation.

The question does not assume replacement of natural-language requirements with their programmatic counterparts: the goal is to have a representation which would encompass the both views with the possibility of extracting only one of them.

\subsubsection{\textbf{RQ2}} \label{sec:sec:sec:rq2}
\textit{How to make each requirement both self-contained and formal?}

Requirements (REQ1)-(REQ11) are already self-contained and thus are suitable for agile development of arbitrary sized increments. How to enrich them with formality without sacrificing their granularity?

\subsubsection{\textbf{RQ3}} \label{sec:sec:sec:rq3}
\textit{How to understand whether the partially available implementation in \autoref{fig:verifiable_clock_implementation} meets requirements (REQ1)-(REQ8)?}

It is possible to take requirements (REQ1)-(REQ11) and mentally convert them directly to a correct implementation, but the task assumes reuse of the existing class \e{CLOCK} in case of its correctness. How can one prove automatically that it meets requirements (REQ1)-(REQ8)? The only available part of the \e{CLOCK} class is its contract - in particular, the postcondition of command \e{tick}. It is also known that the hidden implementation of \e{tick} provably meets its postcondition. The question then reduces to the following one: how can one understand if the postcondition of \e{tick} meets requirements (REQ1)-(REQ8)?

\section{Seamless Requirements} \label{sec:solution}

\begin{figure}
\begin{lstlisting}
  req_1 (clock: CLOCK; current_second: INTEGER)
      -- A clock tick increments current
      -- second if it is smaller than 59.
    require
      modify (clock)
      clock.second < 59
      clock.second = current_second
    do
      clock.tick
    ensure
      clock.second = current_second + 1
    end
\end{lstlisting}
\caption{Requirement (REQ1) in the seamless form} \label{fig:req1_alternative}
\end{figure}

\autoref{fig:req1_alternative} contains the (REQ1) requirement in the form of a seamless requirement - a contracted routine with a natural-language comment\footnote{Comments start with a double hyphen \es{--} in Eiffel}. The comment contains the natural-language representation of (REQ1) in \autoref{fig:natural_clock_requirements}. The routine part, together with the signature and the contract parts, constitutes a proof obligation: "for any object \e{clock} of type \e{CLOCK} and any value \e{current_second} of type \e{INTEGER}, such that \e{clock.second < 59} and \e{clock.second = current_second}, an execution of \e{clock.tick} results in \e{clock.second = current_second + 1}". The \e{modify (clock)} clause in the precondition limits side effects of the \e{tick} routine: the routine is allowed to modify only the target object \e{clock} plus any object owned by \e{clock} \cite{polikarpova2014flexible}. It is possible to submit such a proof obligation to an automatic prover. AutoProof verifier fulfills this role for Eiffel programming language used in this example.

The idea to use non-operational routines with pre- and postconditions for complete specification of programs was first proposed in work \cite{anaumchev2016drivers}. The routines are assumed to be expressed only in terms of their formal arguments. That work introduces a new term "specification drivers" to denote such routines and shows that they are expressive enough to fully capture functional semantics of classes. Since specification drivers are, syntactically speaking, routines, it is possible to comment on them with natural-language statements - the ability to comment on routines is natural for any modern programming language. A seamless requirement consists of two important parts:
\begin{itemize}
  \item Specification driver that captures the formal semantics for the requirement.
  \item Natural-language comment on the specification driver that informally captures the semantics.
\end{itemize}

A specification driver is a contracted routine expressed only in terms of its formal arguments and is understandable to AutoProof as a proof obligation.

The structure of a seamless requirement, together with the properties of specification drivers, answers the questions from \autoref{sec:sec:research_questions} and ensures the core properties of seamless requirements, as the following sections illustrate.

    \subsection{RQ1: understandability to developers} \label{sec:sec:method_stakeholder_reuse}

    Seamless requirements are contracted routines, which are programming language constructs understandable to programmers. Natural-language comments on these routines capture the informal representation of requirements that is understandable to customers. This duality makes a seamless requirement understandable to the two principal groups of stakeholders and semantically connects natural-language requirements to the \e{CLOCK} class, thus answering the RQ1 question from \autoref{sec:sec:sec:rq1}.

    The idea of interweaving natural-language prose with programming language constructs was first proposed by Knuth in \cite{knuth1984literate}. One of the theses of the present work is that it makes sense to use the standard commenting mechanism of the underlying programming language for this purpose.

    \subsection{RQ2: introducing formality into agile development} \label{sec:sec:method_models_reuse}

    As their specification driver components are mathematically precise, seamless requirements do not accumulate ambiguity. Specification drivers are expressed completely in terms of their formal arguments, which is why they are also self-contained. The combination of the two properties benefits agile development with formality and does not interfere with its incrementality.

    \subsection{RQ3: utility for development activities}  \label{sec:sec:method_activities_reuse}

    A seamless requirement is a natural-language statement and, at the same time, is a proof obligation. Consequently, to prove correctness of an implementation with respect to a natural-language requirement is to extend this requirement to the seamless form and then try to prove its proof obligation part. The approach also contributes to the following development activities.

        \subsubsection{Requirements documentation} \label{sec:sec:sec:reqs_documentation}

        A requirements document becomes an auxiliary class in the same namespace with the operational classes. Since seamless requirements are self-contained routines, there is no place for a naming conflict in the event of putting together multiple seamless requirements within a single class.
        \autoref{sec:sec:reqs_documentation} illustrates this concept on the clock example.

        \subsubsection{Specification validation} \label{sec:sec:sec:specs_validation}

        Seamless requirements, being proof-obligations understandable to AutoProof, introduce the notion of proving a requirement. Verification by AutoProof is modular: for example, for proving the \e{req_1} requirement in \autoref{fig:req1_alternative} AutoProof will use only the postcondition of the \e{tick} command. The modularity means that it is possible to verify a program with a hidden implementation with respect to a seamless requirement, when only a contract of the program is available. Section \autoref{sec:sec:specs_validation} illustrates the validation process for the existing \e{CLOCK} class.

        \subsubsection{Specification inference} \label{sec:sec:sec:specs_inference}

        It is possible to use seamless requirements for proof-driven development of programs from scratch. The automatic prover drives the process in this case. To infer a specification from a set of seamless requirements is to equip the operational classes with contracts strong enough to prove the requirements. Once the requirements pass verification by AutoProof, the development process switches to the implementation phase. To infer an implementation from a specification is to implement all the operational classes correctly with respect to their contracts \cite{meyer1988object}. The correctness is proved with the same verifier.

\begin{figure}
\begin{tabular}{m{0.55\textwidth} m{0.52\textwidth}}
\begin{lstlisting}
note explicit: wrapping -- For AutoProof.
deferred class CLOCK_REQUIREMENTS
feature
  -- A clock tick:
  req_1 (clock: CLOCK; current_second: INTEGER)
      -- increments current second if it is
      -- smaller than 59.
    require
      modify (clock)
      clock.second < 59
      clock.second = current_second
    do
      clock.tick
    ensure
      clock.second = current_second + 1
    end
  req_2 (clock: CLOCK)
      -- resets current second to 0 if it
      -- equals 59.
    require
      modify (clock)
      clock.second = 59
    do
      clock.tick
    ensure
      clock.second = 0
    end
  req_3 (clock: CLOCK; current_minute: INTEGER)
      -- increments current minute if the time
      -- is HH:MM:59 for MM smaller than 59
    require
      modify (clock)
      clock.second = 59
      clock.minute < 59
      clock.minute = current_minute
    do
      clock.tick
    ensure
      clock.minute = current_minute + 1
    end
  req_4 (clock: CLOCK)
      -- resets current minute to 0 if it equals
      -- 59 and the current second equals 59.
    require
      modify (clock)
      clock.second = 59
      clock.minute = 59
    do
      clock.tick
    ensure
      clock.minute = 0
    end
\end{lstlisting}
&
\begin{lstlisting}
  req_5 (clock: CLOCK; current_minute: INTEGER)
      -- keeps current minute if  current
      -- second is smaller than 59.
    require
      modify (clock)
      clock.second < 59
      clock.minute = current_minute
    do
      clock.tick
    ensure
      clock.minute = current_minute
    end
  req_6 (clock: CLOCK; current_hour: INTEGER)
      -- increments current hour if the time
      -- is HH:59:59 for HH smaller than 23.
    require
      modify (clock)
      clock.second = 59
      clock.minute = 59
      clock.hour < 23
      clock.hour = current_hour
    do
      clock.tick
    ensure
      clock.hour = current_hour + 1
    end
  req_7 (clock: CLOCK)
      -- resets current hour to 0 if the time
      -- is 23:59:59
    require
      modify (clock)
      clock.second = 59
      clock.minute = 59
      clock.hour = 23
    do
      clock.tick
    ensure
      clock.hour = 0
    end
  req_8 (clock: CLOCK; current_hour: INTEGER)
      -- keeps current hour if current second
      -- is smaller than 59.
    require
      modify (clock)
      clock.second < 59
      clock.hour = current_hour
    do
      clock.tick
    ensure
      clock.hour = current_hour
    end
end
\end{lstlisting}
\end{tabular}
\caption{The seamless requirements document corresponding to (REQ1)-(REQ8)} \label{fig:clock_rd}
\end{figure}

\section{Seamless Requirements in Practice} \label{sec:process}

\begin{figure}
    \fbox{\includegraphics[scale=0.45]{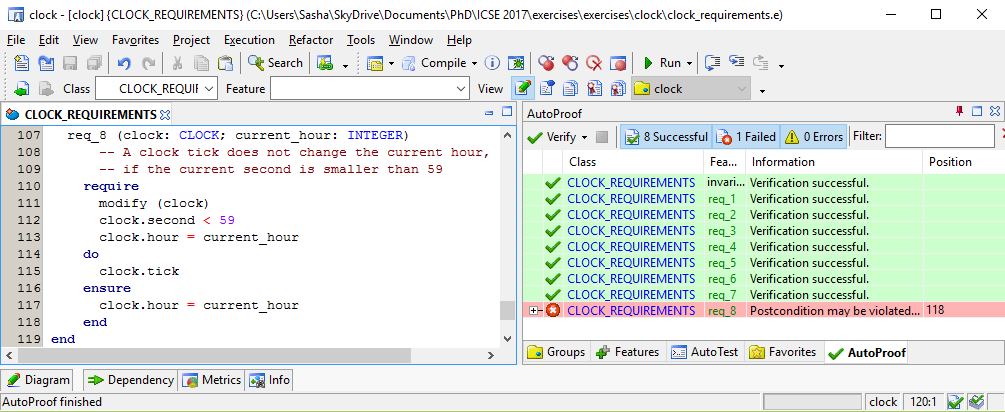}}\\
    \caption{Eiffel Verification Environment with the AutoProof pane}\label{fig:eve_ap}
\end{figure}

\autoref{sec:sec:reqs_documentation} and \autoref{sec:sec:specs_validation} implement step 3 of the development plan from \autoref{sec:sec:clock} by verification of the existing class in \autoref{fig:verifiable_clock_implementation} with respect to requirements (REQ1)-(REQ8). \autoref{sec:sec:zero_increment} and \autoref{sec:sec:incremental} use the verification results as input. The resulting artifacts are publicly available on GitHub \cite{anaumchev2017seamlessrepo}.

\subsection{Requirements documentation} \label{sec:sec:reqs_documentation}

	The first step is to document requirements (REQ1)-(REQ8) in the seamless form. \autoref{fig:clock_rd} contains the respective requirements class\footnote{The ``\es{note explicit: wrapping}" expression at the top of the class is a verification annotation \cite{polikarpova2014flexible} not related to the example.}. The \e{deferred} keyword means that the class is not operational: it is not possible to instantiate any objects from it.

    Since a seamless requirements document is a class, such techniques as inheritance are applicable to it. For example, if a new set of requirements arrives, it is not necessary to add them directly to the \e{CLOCK_REQUIREMENTS} class. It is possible to create a subclass where only new requirements are enumerated, and the old ones will be inherited automatically. \autoref{sec:sec:incremental} illustrates this approach on the clock example.

    The \e{CLOCK_REQUIREMENTS} class is provable by AutoProof: to prove it is to prove each of the seamless requirements it contains. \autoref{sec:sec:specs_validation} describes the meaning of this process.

    \subsection{Specification validation} \label{sec:sec:specs_validation}

    To prove correctness of the \e{CLOCK} class with respect to requirements (REQ1)-(REQ8) is to execute AutoProof on the \e{CLOCK_REQUIREMENTS} class. Verification by AutoProof is modular: verification of a requirements class does not need access to internal implementations of the operational classes, only to their contracts. AutoProof assumes that these implementations meet their respective contracts.

    \autoref{fig:eve_ap} contains a screenshot of Eiffel verification environment (Eve) \cite{tschannen2011usable} with an AutoProof pane on the right side. The AutoProof pane contains the results of verifying the \e{CLOCK_REQUIREMENTS} class. Apparently, the postcondition of the \e{tick} feature in \autoref{fig:verifiable_clock_implementation} is insufficiently strong to meet the (REQ8) requirement. Although the hidden implementation of the \e{CLOCK} class is known to meet its contract, it is possible for the implementation not to meet the requirements. Double-clicking the red line in the AutoProof pane retargets Eve to the \e{req_8} routine, which represents the seamless form of (REQ8).

\begin{figure}
\begin{lstlisting}
class CLOCK
feature
  second, minute, hour: INTEGER

  tick
    do
      -- To implement
    ensure
      -- To specify
    end
end
\end{lstlisting}
\caption{Blank clock implementation of clock} \label{fig:blank_clock_implementation}
\end{figure}

    Since the specification of \e{CLOCK} failed validation with respect to requirements, step 5 of the development plan from \autoref{sec:sec:clock} becomes active. This step consists of developing a completely new \e{CLOCK} class. \autoref{sec:sec:zero_increment} describes development of the regular clock functionality (REQ1)-(REQ8), and \autoref{sec:sec:incremental} incrementally extends it with the days counter functionality (REQ9)-(REQ11). The starting point is a blank class \e{CLOCK} in \autoref{fig:blank_clock_implementation}, which does not have any contract or executable instructions. It only declares the clock features so that the requirements class in \autoref{fig:clock_rd} could compile.

    \subsection{Increment 0: the basic functionality} \label{sec:sec:zero_increment}

	This section describes development of the basic clock functionality increment. The development occurs as follows: once all requirements for the increment are collected, software specification is inferred from them; then, an implementation is inferred to meet the specification. The present section illustrates how application of seamless requirements may facilitate the transitions between the adjacent phases with the help of AutoProof. \autoref{sec:sec:sec:requirements_to_specification} describes inference of a correct \e{CLOCK} specification based on the seamless requirements from the \e{CLOCK_REQUIREMENTS} class. \autoref{sec:sec:sec:specification_to_implementation} infers an implementation of the \e{CLOCK} class that meets the inferred specification.

    \subsubsection{Specification inference} \label{sec:sec:sec:requirements_to_specification}

        \autoref{fig:tick_specification} depicts a postcondition of the \e{tick} feature, which meets the\\
        \e{CLOCK_REQUIREMENTS} class, so that the latter passes verification by AutoProof. How can one infer postconditions from seamless requirements? This problem does not seem to be solvable in the general case; however, the seamless requirements from the\\
        \e{CLOCK_REQUIREMENTS} class possess some common properties:
        \begin{itemize}
          \item Each of them involves only one feature call.
          \item Each of them involves only one object of type \e{CLOCK}.
          \item The \e{tick} feature does not accept any formal arguments.
        \end{itemize}
        These observations enable application of the following inference logic. The resulting assertion takes the form of a logical implication. If a seamless requirement involves some object \e{o: TYPE}, then for every expression of the form \e{o.q}, where \e{q} is a query of class \e{TYPE}, the following rules apply:
        \begin{itemize}
          \item If \e{o.q} occurs in the precondition of the requirement, it translates to \e{old q} in the antecedent of the implication.
          \item If \e{o.q} occurs in the postcondition of the requirement, it translates to \e{q} in the consequent of the implication.
        \end{itemize}
        A requirement may also use an auxiliary formal argument \e{a: SUPPLEMENTARY_TYPE}, such as \e{current_hour: INTEGER} in \e{req_6}. Assume that the following conditions hold together:
        \begin{itemize}
          \item The precondition of the requirement contains an expression of the form \e{o.p = a}.
          \item The postcondition of the requirement contains an expression of the form \e{o.q = f(a)}.
        \end{itemize}
        In this case these conditions translate to \e{q = f(old p)} in the consequent of the resulting implication.

\begin{figure}
\begin{lstlisting}
tick
  do
    -- To implement
  ensure
    old second < 59 implies second = old second + 1
    old second = 59 implies second = 0
    old second = 59 and old minute < 59 implies minute = old minute + 1
    old second = 59 and old minute = 59 implies minute = 0
    old second < 59 implies minute = old minute
    old second = 59 and old minute = 59 and old hour < 23 implies hour = old hour + 1
    old second = 59 and old minute = 59 and old hour = 23 implies hour = 0
    old second < 59 implies hour = old hour
  end
\end{lstlisting}
\caption{Postcondition of \es{tick} that meets \es{req_1}-\es{req_8}} \label{fig:tick_specification}
\end{figure}

        Each assertion from the postcondition in \autoref{fig:tick_specification} is the result of an application of these inference rules to the respective seamless requirement.

        \subsubsection{Implementation inference} \label{sec:sec:sec:specification_to_implementation}

\begin{figure}
\begin{lstlisting}
tick
  do
    if second < 59 then second := second + 1
    else second := 0
      if minute < 59 then minute := minute + 1
      else minute := 0
        if hour < 23 then hour := hour + 1
        else hour := 0
        end
      end
    end
  ensure
    -- Postcondition assertions
  end
\end{lstlisting}
\caption{Implementation of \es{tick} that meets \es{req_1}-\es{req_8}} \label{fig:tick_implementation}
\end{figure}

        Once there is a contract that meets the requirements class, and the latter passes verification by AutoProof, it makes sense to proceed to inference of an implementation that meets the inferred contract. \autoref{fig:tick_implementation} contains an implementation of the \e{tick} feature, which is correct with respect to the postcondition in \autoref{fig:tick_specification}. As in the case of specification inference from requirements, the correctness may be established by an application of AutoProof, but this time it should be executed on the \e{CLOCK} class, which implements the required functionality. The details of the inference process are omitted because they are studied very well \cite{meyer2009touch} and are irrelevant to the central idea of the present article.

	\subsection{Added functionality} \label{sec:sec:incremental}

\begin{figure}
\begin{tabular}{m{0.55\textwidth} m{0.52\textwidth}}
\begin{lstlisting}
note explicit:wrapping
deferred class EXTENDED_CLOCK_REQUIREMENTS
  -- The present class contains requirements
  -- for a clock equipped with a days counter.
inherit CLOCK_REQUIREMENTS
feature
  -- A clock tick:
  req_9 (clock: CLOCK; current_day: INTEGER)
      -- increments current day at 23:59:59,
      -- if it is not Sunday.
    require
      modify (clock)
      clock.second = 59
      clock.minute = 59
      clock.hour = 23
      clock.day < 6
      clock.day = current_day
    do
    	clock.tick
    ensure
    	clock.day = current_day + 1
    end
\end{lstlisting}
&
\begin{lstlisting}
  req_10 (clock: CLOCK)
      -- resets current day to Monday after
      -- a clock tick at 23:59:59 on Sunday.
    require
      modify (clock)
      clock.second = 59
      clock.minute = 59
      clock.hour = 23
      clock.day = 6
    do
      clock.tick
    ensure
      clock.day = 0
    end
  req_11 (clock: CLOCK; current_day: INTEGER)
      -- keeps current day if current
      -- second is smaller than 59.
    require
      modify (clock)
      clock.second < 59
      clock.day = current_day
    do
      clock.tick
    ensure
      clock.day = current_day
    end
end
\end{lstlisting}
\end{tabular}
\caption{The seamless requirements document for extended clock} \label{fig:extended_clock_rd}
\end{figure}

    The regular clock functionality was implemented in \autoref{sec:sec:zero_increment} as one increment. The present section extends the basic functionality in smaller increments consisting of one requirement each.

    There are three requirements in \autoref{sec:problem} that describe the desirable behavior of the clock with a day counter: (REQ9), (REQ10), and (REQ11). \autoref{fig:extended_clock_rd} shows them as a part of a requirements class \e{EXTENDED_CLOCK_REQUIREMENTS}. This class is inherited from the original \e{CLOCK_REQUIREMENTS} class, together with all the existing seamless requirements, to which it adds its own. In the present section, each of the newly added requirements corresponds to a separate increment.

    Compilation of the new requirements class fails: seamless requirements \e{req_9}-\e{req_11} use feature \e{day}, which is not a part of the \e{CLOCK} class yet. To fix the compilation error is to add the respective attribute to the existing list of clock attributes:
\begin{lstlisting}
second, minute, hour, day: INTEGER
\end{lstlisting}
    Now that the new requirements class compiles, it is possible to proceed to the first increment.
        \subsubsection{Increment 1} \label{sec:sec:sec:increment_1}
        Implementation of the first increment starts with submitting the\\
        \e{EXTENDED_CLOCK_REQUIREMENTS} class to formal verification by AutoProof.
        The new seamless requirements \e{req_9}, \e{req_10} and \e{req_11} fail the verification attempt: the postcondition of the \e{tick} command does not say anything about the \e{day} attribute, which has just been added to the operational class. We choose to implement the \e{req_9} requirement in the first increment.

        According to the inference rules from \autoref{sec:sec:sec:requirements_to_specification}, it should suffice to strengthen the postcondition of \e{tick} with the following assertion:
\begin{lstlisting}
old second = 59 and old minute = 59 and old hour = 23 and old day < 6 implies day = old day + 1
\end{lstlisting}
        Now that \e{req_9} passes verification, it is necessary to verify the \e{CLOCK} class. The verification attempt fails because the implementation of \e{tick} has not been updated yet to meet the new assertion in the postcondition.

        The following \e{if} block meets the new assertion, which may be confirmed with AutoProof:

\begin{lstlisting}
tick
  do
    -- Other lines of code
    else hour := 0
      if day < 6 then day := day + 1
      end
    end
  end
\end{lstlisting}
        The new code goes after the existing \e{hour := 0} line: the current day updates only when the current hour resets to 0, meaning at midnight. The first increment is done: AutoProof successfully verifies both the \e{req_9} seamless requirement and the operational class \e{CLOCK}.

        \subsubsection{Increment 2} \label{sec:sec:sec:increment_2}
        Seamless requirements \e{req_10} and \e{req_11} still fail verification of the\\
        \e{EXTENDED_CLOCK_REQUIREMENTS} class.
        The second increment consists of implementing the \e{req_10} requirement. This requirement describes the conditions, under which a clock tick resets the current day to Monday.

        Applying the rules from \autoref{sec:sec:sec:requirements_to_specification} to \e{req_10} results in the following postcondition assertion:
\begin{lstlisting}
old second = 59 and old minute = 59 and old hour = 23 and old day = 6 implies day = 0
\end{lstlisting}
        Correctness of the inferred assertion follows from the fact that \e{req_10} now passes verification by AutoProof.

        Attempts to verify the \e{CLOCK} class fail, which means that the current implementation of the \e{tick} feature does not meet the new postcondition assertion. The antecedent of the assertion is different from the preceding one only in the day-related part. This naturally leads to extending the \e{if} block, introduced in \autoref{sec:sec:sec:increment_1}, with an \e{else} block:
\begin{lstlisting}
tick
  do
    -- Other lines of code
    else hour := 0
      if day < 6 then day := day + 1
      else day := 0
      end
    end
  end
\end{lstlisting}
        An application of AutoProof to the \e{CLOCK} class confirms correctness of the modified implementation.

        \subsubsection{Increment 3} \label{sec:sec:sec:increment_3}

        The last increment consists of implementing the seamless requirement \e{req_11}. The requirement states that nothing happens to the current day in the event of a tick if the current second is smaller than 59.

        Here is the new assertion that results from applying the postcondition inference rules to \e{req_11}:
\begin{lstlisting}
old second < 59 implies day = old day
\end{lstlisting}
        This time not only the seamless requirement passes verification by AutoProof: the existing implementation of the \e{tick} feature does not need any changes, which follows from the fact that the \e{CLOCK} class passes verification. Since the \e{req_11} is, essentially, a safety requirement ("nothing bad happens"), this result should not come as a surprise: no malicious code was introduced during implementation of the preceding requirements.

        Implementation of the new seamless requirements is done: both the requirements class \e{EXTENDED_CLOCK_REQUIREMENTS} and the respective operational class \e{CLOCK} pass verification by AutoProof.

\section{Related Work} \label{sec:related}

	\subsection{Dafny} \label{sec:sec:dafny}

	Dafny \cite{leino2010dafny} is a direct example of a setting other than Eiffel/AutoProof in which the seamless requirements method is directly applicable. The verification approach which AutoProof currently uses is more complicated than that of Dafny (partially because Dafny does not support inheritance and information hiding, but not only), which is why it may make more sense to use the latter for getting familiar with seamless requirements.

    \subsection{Test-Driven Development} \label{sec:sec:tdd}

	Although testing is fundamentally different from program proving, software development through seamless requirements have much in common with test-driven development (TDD) \cite{beck2003test} in terms of the software process. It may be convenient to perceive the new software process as test-driven development where specification drivers replace tests, natural language comments on the specification drivers capture user stories, and program proving replaces testing. One may talk about \textit{\textbf{verification-driven development}} to emphasize these analogies with TDD.

	\subsection{State-based notations} \label{sec:sec:state}
    
	State-based specifications characterize the admissible system states at some arbitrary snapshot \cite{van2009requirements}. Languages such as Z, VDM, B, Alloy, OCL rely on the state-based paradigm. The absence of the imperative layer is what makes state-based notations inapplicable for specification of abstract requirements. State-based notations are purely declarative notations in which one cannot say ``if some property holds for a set of objects and I modify some of them through some commands, then another property will hold for these objects''.

	\subsection{Goal-Oriented Requirements Engineering} \label{sec:sec:goal_oriented_re}

\begin{figure}
	\textbf{Goal} Maintain[TrackSegmentSpeedLimit]\\	
	\hspace*{2mm}\textbf{InformalDef} \textit{A train should stay below the maximum speed the track segment can}\\
	\hspace*{21mm}\textit{handle}\\
	\hspace*{2mm}\textbf{FormalDef} $\forall tr: Train, s: TrackSegment \bullet On(tr,s) \implies tr.Speed \leq s.SpeedLimit$
\caption{An example of a goal-oriented requirement from \cite{van2001goal}}
\label{fig:goal_example}
\end{figure}
Goal-oriented requirements \cite{van2001goal} are suitable for addressing the gap between agile and formal development (\autoref{sec:intro}): goals are self-contained and have place for both formal and informal semantics of requirements.

Goals, however, do not bridge the semantical gap between formal requirements notations and programs because the approach does not treat a programming language as a formal notation. Goals also fail to bridge the gap between construction and verification: the need to translate them into tests is still there.
\begin{figure}
\begin{lstlisting}
maintain_track_segment_speed_limit (tr: TRAIN; s: TRACK_SEGMENT)
    -- A train should stay below the maximum speed the track segment can handle
  require
    tr.on (s)
  do
  ensure
    tr.speed <= s.speed_limit
  end	
\end{lstlisting}
\caption{The goal-oriented requirement Maintain[TrackSegmentSpeedLimit] (\autoref{fig:goal_example}) in the form of a seamless requirement.}
\label{fig:seamless_goal_example}
\end{figure}

\begin{figure}
\begin{lstlisting}
maintain_track_segment_speed_limit_without_contract (tr: TRAIN; s: TRACK_SEGMENT)
    -- A train should stay below the maximum speed the track segment can handle
  do
    if tr.on (s) then
      check tr.speed <= s.speed_limit end
    end
  end	
\end{lstlisting}
\caption{The goal-oriented requirement Maintain[TrackSegmentSpeedLimit] (\autoref{fig:seamless_goal_example}) in the form of a seamless requirement without a contract.}
\label{fig:seamless_goal_example_no_contract}
\end{figure}

Seamless requirements approach, while bringing the same benefits as goals do, offers strong pairwise connection between requirements, specifications and code. The \e{maintain_track_segment_speed_limit} seamless requirement (\autoref{fig:seamless_goal_example}) captures the semantics of the corresponding goal (\autoref{fig:goal_example}) in terms of Eiffel programming constructs understandable to Eiffel programmers, though it may be rewritten without contracts at all through \e{if} and \e{check} (known as ``assert" in other languages) statements (\autoref{fig:seamless_goal_example_no_contract}). The last option may be useful in languages without contracts.
\begin{figure}
\begin{tabular}{l r}
\begin{lstlisting}
class TRAIN
feature
  speed: INTEGER

  on (s: TRACK_SEGMENT): BOOLEAN
    do
    ensure
      Result implies speed <= s.speed_limit
    end
end
\end{lstlisting}
&
\begin{lstlisting}
      class TRACK_SEGMENT
      feature
        speed_limit: INTEGER
      end
\end{lstlisting}
\end{tabular}
\caption{Specification of classes \e{TRAIN} and \e{TRACK_SEGMENT} that meets the \e{maintain_track_segment_speed_limit} requirement (\autoref{fig:seamless_goal_example}).} \label{fig:strong_train_spec}
\end{figure}
Successful verification of the \e{maintain_track_segment_speed_limit} requirement assumes strong enough specification of classes \e{TRAIN} and \e{TRACK_SEGMENT} (\autoref{fig:strong_train_spec}). Successful verification of the specified classes assumes, in its turn, implementing the \e{TRAIN::on} routine correctly.

    \subsection{Literate Programming} \label{sec:sec:literate}

    Knuth in was the first one to apply interwoven notations in programming \cite{knuth1984literate}. Meyer criticized the approach as inapplicable to object-oriented programming and proposed the multirequirements \cite{Meyer13Multi} method (\autoref{sec:sec:multirequirements}):
    \begin{quote}
    When I first read about literate programming I was seduced by the elegance of the approach, but found it inapplicable to modern, object-oriented programming which (as discussed in several publications including \cite{meyer1988object}) is fundamentally bottom-up as implied by the focus on reuse; literate programming seemed inextricably tied to the top-down, function-driven programming style of the nineteen-seventies. In that traditional view, a program implements a single "main" function; as a consequence the "literate" text is the sequential telling, cradle to grave, of a single story.
    \end{quote}

    \subsection{Multirequirements} \label{sec:sec:multirequirements}

    A multirequirement is a combination of a natural-language statement and a small piece of the resulting program; the program piece should rephrase what the natural-language part says. The multirequirements method \cite{Meyer13Multi} adapts Knuth's idea of interwoven notations to object-oriented programming, while focusing on traceability. The method suggests using three notation layers: natural-language layer, formal layer, and graphical layer. For the formal layer, it suggests usage of pieces of the presumable final program. When the requirements specification phase is over, specialized tools then take those pieces and merge them into the program skeleton. The tools are also responsible for taking care of both up- and down-traceability. The approach conceptually removes the fundamental flaw of literate programming, which is the need to write a complete story from the beginning to the end.

    Michael Jackson in his work \cite{jackson2014topsy} extensively criticizes piecemeal construction of cyber-physical systems. Apart from the details of that particular work, the multirequirements method possesses several flaws that are of concern for us:
    \begin{itemize}
      \item The presumed additional tools responsible for keeping the requirements document and the resulting program in sync do not seem easy to implement. The method assumes that any person responsible for requirements specification admits the concern for traceability and connects natural-language descriptions with the corresponding program pieces through special anchors. As a consequence, the tools should also be able to detect mistakenly placed anchors as well as to warn of their potential absence.
      \item The method is applicable only to "forward" development. There is no way to prove that an existing program meets a multirequirement. The programmatic part of a multirequirement is, conceptually, a small piece of the program itself. In order to submit a multirequirement to formal verification, it is necessary first to integrate that piece into the main program. This process changes the original program, which is why the very notion of verifying a program with respect to a multirequirement does not exist.
      \item The multirequirements method assumes a strong bias of the requirements specification notation toward features of a specific programming language (Eiffel in that particular work). A seamless requirement is simply a command with a pre- and a postcondition expressed only in terms of its formal arguments. Such commands are a kind of construct available in any modern programming language with contracts, such as Dafny, Spec\# or D.
    \end{itemize}

Applicability of the multirequirements method has also been studied on a realistic example \cite{naumchev2015unifying}.

\section{Summary} \label{sec:summary}

As the development case illustrates, seamless requirements empower software engineering with the following properties:

\begin{itemize}
  \item Unity of software construction and verification: seamless requirements stimulate construction and, at the same time, are suitable for checking correctness of the deliverables.
  \item Unity of functional requirements and code: the requirements document becomes one of the classes in the source code repository, readable by both customers and developers.
  \item Independence from a particular development model choice: there is no need to adjust the requirements notation in the event of switching the development model on the go.
  \item Traceability for free: existing features of the underlying IDE are suitable for traceability management in the following form:
    \begin{itemize}
      \item to trace a seamless requirement to specification (downward traceability \cite{van2009requirements}) is to retarget the IDE to the definitions of the operational classes and features that occur in the requirement; this functionality is present in some form in any modern IDE.
      \item tracing a class or a feature to requirements that constrain it (upward traceability \cite{van2009requirements}) reduces to an application of the "Show Callers" feature, which is also present in all modern IDE's (up to a name); every call from the requirements class is done by some seamless requirement.
    \end{itemize}
\end{itemize}

    \subsection{Limitations of the example} \label{sec:sec:example_limitation}

    Several potential complications were ignored in favour of simplicity of the narrative:

    \begin{itemize}
      \item There is only one command in the clock example: \e{tick}. Despite this, the approach scales to multi-command examples. As work \cite{anaumchev2016drivers} demonstrates, specification drivers, which serve as the formal layer of seamless requirements, are capable of handling cases with an arbitrary number of commands.
      \item The \e{tick} command does not accept any formal arguments. In fact, the approach scales to the case with formal arguments: if a seamless requirement describes desirable behavior of a command with a formal argument, the corresponding routine may assume the presence of the argument through extending the list of its own formal arguments. Work \cite{anaumchev2016drivers} uses this technique too.
    \end{itemize}

    The postcondition inference logic from \autoref{sec:sec:sec:requirements_to_specification} only work in the context of these two simplifications. In general, inference of a sufficiently strong postcondition does not seem to be a solvable problem.

    \subsection{Limitations of the approach} \label{sec:sec:approach_limitation}

    As the primary concern of the work is functional correctness, all questions related to the suitability of seamless requirements for non-functional requirements lie expressly outside of the scope of this paper.

    Another assumption that underlies this approach is the use of a programming language with contracts plus the existence of a prover for this language. This assumption is adequate: Eiffel plus AutoProof is not the only representative of this technology combination. The "Code Contracts for .NET" project \cite{codecontracts} offers similar benefits in the .NET world.

	Seamless requirements approach is applicable to non-concurrent programs. Although the approach may have potential in concurrent setting too, the question is not studied yet.
    \subsection{Future work}

    \subsubsection{Translation between the notations} \label{sec:sec:sec:translator}    
    The seamless requirements approach poses an immediate question: how to check the consistency between the natural-language and the programming language components? Currently there is no way to do that. Work \cite{meyer1993formalism} describes a requirements refinement process that relies on round trip engineering: given a natural-language requirement translate it into a formal form and then back and see how close the result is to the original statement. This process needs support in the form of two tools that would perform the necessary translations. Development of these tools is the immediate goal of the present research.

    \subsubsection{Consistency of seamless requirements} \label{sec:sec:sec:consistency}
    
    Another research question is: how to understand if seamless requirements are consistent with each other? With an inconsistent set of requirements it will never be possible to develop a provably correct solution. With trial-and-error considerable amount of resources may be spent before the inconsistency becomes apparent. How could one detect inconsistencies in requirements before initiating implementation of a solution?

	\subsection*{Acknowledgement}
	The authors are thankful to the administration of Innopolis University for the funding that made this work possible.



%
{{{
\bibliographystyle{ieeetr}
\bibliography{SeamlessRequirements}

\begin{thebibliography}{10}

\bibitem{knuth1984literate}
D.~E. Knuth, ``Literate programming,'' {\em The Computer Journal}, vol.~27,
  no.~2, pp.~97--111, 1984.

\bibitem{Meyer13Multi}
B.~Meyer, ``Multirequirements,'' in {\em Modelling and Quality in Requirements
  Engineering (Martin Glinz Festscrhift)} (N.~Seyff and A.~Koziolek, eds.), MV
  Wissenschaft, 2013.

\bibitem{tschannen2015autoproof}
J.~Tschannen, C.~A. Furia, M.~Nordio, and N.~Polikarpova, ``Autoproof:
  Auto-active functional verification of object-oriented programs,'' {\em arXiv
  preprint arXiv:1501.03063}, 2015.

\bibitem{van2009requirements}
A.~Van~Lamsweerde {\em et~al.}, {\em Requirements engineering: from system
  goals to UML models to software specifications}.
\newblock 2009.

\bibitem{anaumchev2016drivers}
A.~Naumchev and B.~Meyer, ``Complete contracts through specification drivers,''
  in {\em 2016 10th International Symposium on Theoretical Aspects of Software
  Engineering (TASE)}, pp.~160--167, July 2016.

\bibitem{van2001goal}
A.~Van~Lamsweerde, ``Goal-oriented requirements engineering: A guided tour,''
  in {\em Requirements Engineering, 2001. Proceedings. Fifth IEEE International
  Symposium on}, pp.~249--262, IEEE, 2001.

\bibitem{turk2014limitations}
D.~Turk, R.~France, and B.~Rumpe, ``Limitations of agile software processes,''
  {\em arXiv preprint arXiv:1409.6600}, 2014.

\bibitem{black2009formal}
S.~Black, P.~P. Boca, J.~P. Bowen, J.~Gorman, and M.~Hinchey, ``Formal versus
  agile: Survival of the fittest,'' {\em Computer}, vol.~42, no.~9, 2009.

\bibitem{drobka2004piloting}
J.~Drobka, D.~Noftz, and R.~Raghu, ``Piloting xp on four mission-critical
  projects,'' {\em IEEE software}, vol.~21, no.~6, pp.~70--75, 2004.

\bibitem{sidky2007determining}
A.~Sidky and J.~Arthur, ``Determining the applicability of agile practices to
  mission and life-critical systems,'' in {\em Software Engineering Workshop,
  2007. SEW 2007. 31st IEEE}, pp.~3--12, IEEE, 2007.

\bibitem{jorgensen2016software}
P.~C. Jorgensen, {\em Software testing: a craftsman’s approach}.
\newblock CRC press, 2016.

\bibitem{meyer1988eiffel}
B.~Meyer, ``Eiffel: A language and environment for software engineering,'' {\em
  Journal of Systems and Software}, vol.~8, no.~3, pp.~199--246, 1988.

\bibitem{meyer1993formalism}
B.~Meyer, ``On formalism in specifications,'' in {\em Program Verification},
  pp.~155--189, Springer, 1993.

\bibitem{polikarpova2014flexible}
N.~Polikarpova, J.~Tschannen, C.~A. Furia, and B.~Meyer, ``Flexible invariants
  through semantic collaboration,'' in {\em FM 2014: Formal Methods},
  pp.~514--530, Springer, 2014.

\bibitem{meyer1988object}
B.~Meyer, {\em Object-oriented software construction}, vol.~2.
\newblock Prentice hall New York, 1988.

\bibitem{anaumchev2017seamlessrepo}
A.~Naumchev, ``Seamless requirements example..''
  \url{https://github.com/anaumchev/seamless_requirements}, 2017.

\bibitem{tschannen2011usable}
J.~Tschannen, C.~A. Furia, M.~Nordio, and B.~Meyer, ``Usable verification of
  object-oriented programs by combining static and dynamic techniques,'' in
  {\em International Conference on Software Engineering and Formal Methods},
  pp.~382--398, Springer, 2011.

\bibitem{meyer2009touch}
B.~Meyer, {\em Touch of Class: learning to program well with objects and
  contracts}.
\newblock Springer, 2009.

\bibitem{leino2010dafny}
K.~R.~M. Leino, ``Dafny: An automatic program verifier for functional
  correctness,'' in {\em International Conference on Logic for Programming
  Artificial Intelligence and Reasoning}, pp.~348--370, Springer, 2010.

\bibitem{beck2003test}
K.~Beck, {\em Test-driven development: by example}.
\newblock Addison-Wesley Professional, 2003.

\bibitem{jackson2014topsy}
M.~Jackson, ``Topsy-turvy requirements,'' {\em Requirements Engineering},
  vol.~19, no.~1, pp.~107--111, 2014.

\bibitem{naumchev2015unifying}
A.~Naumchev, B.~Meyer, and V.~Rivera, ``Unifying requirements and code: an
  example,'' in {\em International Andrei Ershov Memorial Conference on
  Perspectives of System Informatics}, pp.~233--244, Springer, 2015.

\bibitem{codecontracts}
Microsoft, ``Code csontracts for .net.''
  \url{https://visualstudiogallery.msdn.microsoft.com/1ec7db13-3363-46c9-851f-1ce455f66970},
  2015.

\end{thebibliography}
}}}

\end{document}